\documentstyle[aps,epsf,preprint]{revtex}

\begin{document}

\draft

\tighten





\title{Constraints on vector mesons with finite momentum in
nuclear matter}

\author{Bengt Friman$^{1,2}$,
Su Houng Lee $^{2,3}$\footnote{Alexander von Humboldt fellow} and
Hungchong Kim$^{4}$}
\address{$^1$ Institute fur Kernphysik, TU Darmstadt, D-64289 Darmstadt,
Germany \\
$^2$  GSI, Planckstr. 1, D-64291 Darmstadt, Germany \\
$^3$ Department of Physics, Yonsei University, Seoul, 120-749,
Korea  \\
$^4$ Department of Physics, Tokyo Institute of Technology,
Tokyo 152-8551, Japan }


\maketitle

\begin{abstract}
Using the QCD operator product expansion, we  derive the real part
of the transverse and longitudinal vector-vector correlation
function with the quantum numbers of the $\rho$ and $\omega$ mesons
to leading order in density and three momentum (${\bf q}^2$) for
$\omega^2 \rightarrow -\infty $. The operator product expansion
provides, through the Borel transformed energy  dispersion
relation, a model independent constraint for the momentum
dependence of the vector meson spectral density in nuclear matter.
Existing model calculations for the dispersion effect of the
$\rho$, where the vector-meson nucleon scattering amplitude is
obtained by resonance saturation in the s-channel, in general
violate this constraint. We trace this to an inconsistent choice
for the form factor of the $\Delta N\rho$ vertex. With a consistent
choice, where both the form factor and the coupling constant are
obtained from the Bonn potential, the contribution of the $\Delta$
is substantially reduced and we find good agreement with the
constraint equation. We briefly comment on the implications of our
result for attempts to interpret the enhancement of low-mass
dileptons in heavy-ion collisions.

\end{abstract}


\section{INTRODUCTION}

It is expected that the properties of vector mesons at finite
density and/or temperature should be reflected in the dilepton
spectrum observed in nuclear reactions. It has also been suggested
that modifications of the vector-meson properties are related to
the restoration of chiral symmetry in dense
matter~\cite{P82,Lee98}. Indeed the enhancement of low-mass
dileptons found in heavy-ion collisions~\cite{CERES} can be
interpreted in terms of a non-trivial change of the vector meson
spectral density in hot/dense matter. However, the dilepton spectra
observed in heavy-ion collisions involve dileptons from different
stages of the collision. Consequently collision models have to be
invoked to relate the spectra to the vector-meson spectral density
at a given temperature and density. At present, the data can be
qualitatively interpreted in terms of two scenarios; on the one
hand, a scenario where the vector meson masses decrease in
matter~\cite{LKB95,BR91,HL92,Walecka,ST95,FS96} and on the other
hand, a scenario based on hadronic models where the $\rho$-meson
spectrum exhibits a non-trivial dispersion effect (three momentum
dependence)~\cite{FP97,RCW97} and an enhanced
width~\cite{KKW97,KW98}.

In spite of attempts to unify the theoretical descriptions of the
vector meson spectral densities in matter~\cite{BLRRW98}, the
characteristics of the two approaches still differ and the relation
between them remains nebulous. In order to reduce the discrepancies
and eventually approach a unified description, it is important to
reduce the uncertainties of the theoretical models by constraining
the parameters with experimental data~\cite{Friman98,RUBW97} and
results derived directly from QCD.

In this work, we derive in detail QCD constraints on the
three-momentum dependence of the vector meson spectral density in
nuclear matter.  These constraints have been derived previously in
the context of QCD sum rules~\cite{Lee98c,LK98} and shown to imply
a small three-momentum dependence of the vector meson mass at
finite density. Here, we give a detailed derivation of the
constraints and test to what extent it is satisfied by existing
hadronic model calculations. In the first stage we derive the
Operator Product Expansion (OPE) and constructs constraints on the
spectral density using the Borel transformed dispersion relation.
In the second stage, model calculations are confronted with these
constraints. As we will see, the relevant condensates that appear
in the constraints for the dispersion effect are well known up to
nuclear matter density and the corresponding QCD sum rule provides
a solid reference point for model
calculations~\cite{Lee98c,LK98,LM98}.

An alternative approach is to model the phenomenological side by a
simple functional form with a few parameters and obtain medium
modifications of these parameters and thus of the spectral density.
This procedure is in general not unique unless the form of the
spectral density can be constrained by independent experimental or
theoretical input. For the $\rho$ meson, for instance, there is a
trade-off between a reduction of the mass and an increase of the
width. Thus, there is a band of possible combinations of mass and
width, where the quality of the fit to the constraint is almost
unchanged. Furthermore, there are uncertainties related to the
medium expectation value of the four-quark operator, for which the
ground state saturation hypothesis was used. Consequently, the
resulting constraint~\cite{HL92} can be satisfied either by a large
decrease in mass and a small increase in width or by a small
decrease in mass and a large increase in width or something in
between~\cite{KKW97,LPM98}.

The paper is organized as follows.  In section II, we derive the
OPE relevant for the dispersion effect of the vector meson spectral
density. In section III, we state the constraint, which should be
satisfied by model calculations. Finally in section IV, we discuss
to what extent the existing hadronic model calculations satisfy
this constraint.

\section{Operator Product Expansion}

Consider the vector-current correlation function in nuclear matter

\begin{eqnarray}
\Pi_{\mu\nu} (\omega, {\bf q} ) &=& i \int d^4x e^{iq\cdot x}\langle G|
      T [ J_\mu(x) J_\nu(0) ] |G \rangle\ ,
\label{ope1}
\end{eqnarray}
where $J_\mu=\bar{q} \gamma_\mu q$, $ |G \rangle$ is the nuclear
ground state at rest and $q^\mu =(\omega,{\bf q})$. In what
follows, when we give results for explicit vector mesons, we will
use the currents $J_\mu^{\rho,\omega}=\frac{1}{2} ( {\bar u}
\gamma_\mu u \mp {\bar d} \gamma_\mu d )$ for the $\rho$ and
$\omega$ mesons, respectively.

In general, because the vector current is conserved, the
polarization tensor in  eq.~(\ref{ope1}) has only two independent
functions~\cite{tensors}

\begin{eqnarray}
\Pi_{\mu\nu}(\omega,{\bf q})=\Pi_T q^2 {\rm P}^T_{\mu\nu}+ \Pi_L q^2
 {\rm P}^L_{\mu\nu}\ ,
\label{ope2}
\end{eqnarray}
where we assume the nuclear medium to be at rest, such that $ {\rm
P}^T_{00} =  {\rm P}^T_{0i}={\rm P}^T_{i0}=0$, ${\rm P}^T_{ij}  =
\delta_{ij}-{\bf q}_i {\bf q}_j/{\bf q}^2 $ and $ {\rm
P}^L_{\mu\nu}  =  (q_\mu q_\nu/q^2-g_{\mu\nu}- {\rm
P}^T_{\mu\nu})$. The longitudinal and transverse polarization
functions are given by
\begin{eqnarray}
\Pi_L=\frac{1}{{\bf q}^2} \Pi_{00}\ ,~~~~~ \Pi_T=-\frac{1}{2}
(\frac{1}{q^2} \Pi_{\mu}^{\mu}+ \frac{1}{{\bf q}^2} \Pi_{00})\ .
\label{ope3}
\end{eqnarray}
In the limit of ${\bf q} \rightarrow 0$, the longitudinal and
transverse polarizations are degenerate, i.e., $\Pi_L=\Pi_T$.

In this work, we address only the leading three momentum (${\bf
q}$) dependence of the polarization functions. Consequently, we
expand the correlation function in a Taylor series around ${\bf q}
= 0$ and examine the dispersion relation at fixed ${\bf q}$,
\begin{eqnarray}
{\rm Re} \Pi_{L,T}(\omega^2,{\bf q}^2)  =  {\rm Re} \left( \Pi^0(\omega^2,0)+
\Pi_{L,T}^1(\omega^2,0) ~ {\bf q}^2 + \cdot \cdot \cdot \right) \nonumber \\
  =  \int_0^\infty du^2 \left( {\rho_{L,T}^0(u,0) \over (u^2-\omega^2)}
+ {\rho_{L,T}^1(u,0) \over (u^2-\omega^2)} ~ {\bf q}^2 + \cdot
\cdot \cdot
\right)\ ,
\label{ope4}
\end{eqnarray}
where $\rho(u,{\bf q})=(1/\pi) {\rm Im} \Pi^R(u ,{\bf q})$,  and
$R$ denotes the retarded correlation function. The OPE for $\Pi^0$
appropriately modified for the nuclear medium together with the
corresponding energy dispersion relation provides constraints on
the mass shift and spectral changes at ${\bf q}=0$~\cite{HL92}. In
this work, we will explore the three momentum dependence by
studying the OPE in medium for $\Pi_{L,T}^1$.

In general, the OPE~\cite{Wilson69,muta} for the polarization
function at large $Q^2=-\omega^2+{\bf q}^2$ is of the form,
\begin{eqnarray}
\label{ope5}
\Pi_{\mu \nu}(\omega,{\bf q}) &  = &
(q_\mu q_\nu-g_{\mu \nu} q^2)
\left[ -c_0 {\rm ln}|Q^2|+ \sum_n {1 \over Q^n} A^{n,n} \right]
\nonumber \\[10pt]
& & + \sum_{\tau=2} \sum_{k=1} [-g_{\mu \nu} q_{\mu_1} q_{\mu_2}
+g_{\mu \mu_1} q_\nu q_{\mu_2}+q_\mu q_{\mu_1}g_{\nu \mu_2} +
g_{\mu \mu_1} g_{\nu \mu_2} Q^2] \nonumber \\
 &\times & q_{\mu_3} \cdot \cdot \cdot  q_{\mu_{2k}}
\frac{2^{2k}}{Q^{4k+\tau-2}} A^{2k+\tau,\tau}_{\mu_1 \cdot \cdot \cdot \mu_{2k}}
\nonumber \\[10pt]
& & + \sum_{\tau=2} \sum_{k=1} [g_{\mu \nu}- q_\mu q_\nu/q^2]
q_{\mu_1} \cdot \cdot \cdot q_{\mu_{2k}}
\frac{2^{2k}}{Q^{4k+\tau-2}} C^{2k+\tau,\tau}_{\mu_1 \cdot \cdot \cdot
\mu_{2k}}\ .
\end{eqnarray}
Here we have extracted the trivial $\frac{q^\alpha \cdot \cdot}{
Q^n}$ dependence so that $A^{d,\tau},C^{d,\tau}$ represent the
residual Wilson coefficient times  matrix element of an operator of
dimension $d$ and twist $\tau=d-s$, where $s=2k$ is the number of
spin indices of the operator.  The first set of terms is due to the
OPE of scalar operators, while the second set originates from
operators with spin, which have been written as a double sum in
twist and spin. The coefficients $A,C$ represent the two linearly
independent sums of operators, which reflect the two polarization
directions.

The longitudinal and transverse polarization functions can be
obtained by using eq.~(\ref{ope5}) in eq.~(\ref{ope3}). The ${\bf
q}$ dependence emerging from the first line of  eq.~(\ref{ope5}),
namely the contributions from the scalar operators,  is hidden in
$Q^2$. This we call the ``trivial'' ${\bf q}$ dependence, where the
momentum ${\bf q}$ appears in the Lorenz invariant combination
$\omega^2-{\bf q}^2$. Here, we are interested only in the
``non-trivial'' ${\bf q}$ dependence of $\Pi_{L,T}^1$. Consequently
the scalar operators do not contribute to $\Pi_{L,T}^1$ and only
operators with spin contribute to the non-trivial ${\bf q}$
dependence. This is done by expressing  the polarization function
in terms of $Q^2$ and ${\bf q}^2$, i.e., $\Pi(Q^2, {\bf q}^2)$ and
extracting the term linear in ${\bf q}^2$.

\subsection{Linear density approximation}

To leading order in density
\begin{eqnarray}
\label{ope7}
\langle G| A | G \rangle =\langle 0| A | 0 \rangle
 + \frac{\rho_n}{2 m} \langle p| A | p\rangle \ ,
\end{eqnarray}
where $| G \rangle$ denotes the ground state of nuclear matter. The
first term on the right hand side is the vacuum expectation value,
which vanishes for operators with spin. In the second term
$|p\rangle $ denotes a nucleon state with the normalization
$\langle p|  | p^\prime\rangle=2p_0
\delta^3(p-p^\prime)$, $\rho_n$ the nuclear density
and $m$ the nucleon mass. As in vacuum, we include operators up to
dimension 6 in the OPE. This implies that in eq.~(\ref{ope5}) there
are contributions from $(\tau,s)=(2,2),(2,4)$ and $(4,2)$. The
nucleon matrix elements of the $\tau=2$ operators are very well
determined. The $\tau=4$ matrix element appearing in the
$\rho,\omega$ sum rule are similar to those appearing in deep
inelastic scattering (DIS) of electrons off nucleons~\cite{Jaffe}.
These matrix elements have been estimated~\cite{CHKL93,L94} with an
uncertainty of about $\pm$ 30\% using DIS data from CERN and SLAC.

The operators with spin indices are symmetric and traceless so that
when the matrix elements are taken, the general form can be written
as follows~\cite{Nacht73,GP76},
\begin{eqnarray}
\label{ope8}
A_{\mu_1 \cdot \cdot \cdot \mu_{2k}}= \left(
 \sum_{j=0}^k a_j p^{2j} [ g_{\mu_1 \mu_2}\cdot \cdot \cdot g_{\mu_{2j-1}
\mu_{2j}}
   p_{\mu_{2j+1}}
\cdot  \cdot \cdot p_{\mu_{2k}} + permutations ] \right) A_{2k} \ ,
\end{eqnarray}
where there are $\frac{(2k)!}{(2k-2j)!2^j j!}$ terms for a given
value of $k$ and $j$ and $a_j=(-1)^j \frac{(2k-j)!}{2^j (2k)!}$
(more details are given in Appendix A). For $s=2$ and $s=4$ one
finds
\begin{eqnarray}
\label{ope9}
A_{\mu \nu} & = & (p_\mu p_\nu-\frac{1}{4} g_{\mu \nu} p^2) A_2\ , \nonumber \\
A_{\mu \nu \alpha \beta}
& = &
\biggl[p_\mu p_\nu p_\alpha p_\beta -\frac{p^2}{8}
(p_\mu p_\nu g_{\alpha \beta}+p_\mu p_\alpha g_{\nu \beta}
+p_\mu p_\beta g_{\alpha \nu}+p_\nu p_\alpha g_{\mu \beta}
+p_\nu p_\beta g_{\mu \alpha}+p_\alpha p_\beta g_{\mu \nu}) \nonumber
\\[12pt]
& &   ~~~~~~~~~~~~~~~~+\frac{p^4}{48} (g_{\mu \nu} g_{\alpha \beta}+
g_{\mu \alpha} g_{\nu \beta}+
g_{\mu \beta} g_{\nu \alpha}) \biggl] A_4\ .
\end{eqnarray}
Substituting eq.~(\ref{ope8})  into eq.~(\ref{ope5}), we obtain the
contributions of the spin dependent operators to linear order in
density
%
\begin{eqnarray}
\label{ope10}
\Pi_{\mu \nu}&=  & \frac{\rho_n}{2m}\left[ d_{\mu \nu} \sum_{\tau=2} \sum_{k=1}
{ A_{2k}^{2k+\tau,\tau} \over Q^{\tau-2}}
 \sum_{j=0}^{k}  \left( \frac{m^2}{Q^2}
\right)^j \frac{1}{x^{2k-2j}} { (2k-j)! \over (2k) (2k-1) (2k-2j-2)!
j! } \right. \nonumber \\[12pt]
&+ & e_{\mu \nu} \sum_{\tau=2} \sum_{k=1}
 { C_{2k}^{2k+\tau,\tau}\over Q^{\tau-2}}
 \sum_{j=0}^{k}  \left( \frac{m^2}{Q^2}
\right)^j \frac{1}{x^{2k-2j}} { (2k-j)! \over (2k-2j)!
j! }  \nonumber \\[12pt]
&- & \left. e_{\mu \nu}
\sum_{\tau=2} \sum_{k=1} { A_{2k}^{2k+\tau,\tau} \over Q^{\tau-2}}
\sum_{j=1}^{k}  \left( \frac{m^2}{Q^2}
\right)^j \frac{1}{x^{2k-2j}} {4 \cdot (2k-j) \cdot
 (2k-j)! \over (2k) (2k-1) (2k-2j)! (j-1)! }\right]\ ,
\end{eqnarray}
where $x=\frac{Q^2}{2 p \cdot q}$,
 $ e_{\mu \nu}=g_{\mu \nu}-q_\mu q_\nu/q^2$ and
 $d_{\mu \nu}=-p_\mu p_\nu q^2/(p \cdot q)^2+(p_\mu q_\nu+p_\nu
q_\mu)/ p \cdot q -g_{\mu \nu}$. The latter two are the usual
polarization tensors used in DIS~\cite{muta}. The sum over $k$
corresponds to a sum over the number of spin indices while the sum
over $j$ is a sum over the target mass corrections. It should be
noted that in the Bjorken limit $x
\rightarrow~ finite$ , $Q^2 \rightarrow~ \infty$, the leading terms
are twist-2 with any number of spins, while the higher twist terms
and the target mass corrections are suppressed by powers of
${1}/{Q^2}$.

The  nontrivial contribution to $\Pi_{L,T}^1$ can be obtained as
follows. First, we use eq.~(\ref{ope10}) in eq.~(\ref{ope3}). Since
we have assumed the  ground state to be at rest, the nucleons are
also at rest $p=(m,0)$ so that $e_{00}=-\frac{{\bf q}^2}{q^2},
d_{00}=\frac{{\bf q}^2}{\omega^2}, e_\mu^\mu=3,
d_\mu^\mu=-\frac{q^2}{\omega^2}-2$. We then extract the term
proportional to ${\bf q}^2$ as discussed above. For the $s=2$ terms
we find
\begin{eqnarray}
\Pi^1_L  & = & \frac{\rho_n}{2m} \sum_\tau (-1)^{\tau/2}
\frac{ m^2}{\omega^{\tau+4}}
 (4C^{2+\tau,\tau}_2)\ , \nonumber  \\
\Pi^1_T  & = &\frac{\rho_n}{2m} \sum_\tau (-1)^{\tau/2} \frac{ m^2}{\omega^{\tau+4}}
 ( 4C^{2+\tau,\tau}_2-4A^{2+\tau,\tau}_2 )\ ,
\label{ope11}
\end{eqnarray}
while for $s=4$,
\begin{eqnarray}
\Pi^1_L  & = & \frac{\rho_n}{2m} \sum_\tau (-1)^{\tau/2+1} \frac{ m^4}{\omega^{\tau+6}}
 (4A^{4+\tau,\tau}_4-20C^{4+\tau,\tau}_4 )\ , \nonumber  \\
\Pi^1_T  & = &\frac{\rho_n}{2m} \sum_\tau (-1)^{\tau/2+1} \frac{ m^4}{\omega^{\tau+6}}
 (18A^{4+\tau,\tau}_4- 20C^{4+\tau.\tau}_4 )\ .
\label{ope12}
\end{eqnarray}
In general the contribution of operators with $s$ spin indices is a
sum of terms of order  ${1}/{\omega^{\tau+s+2}}$.

\subsection{Form of $A,C$ for $\tau=2$ in the linear density approximation}

The coefficients $A$ and $C$ can be determined by comparing
eq.~(\ref{ope10}), in the Bjorken limit, with the expectation value
of the polarization tensor (\ref{ope5}) in a one-nucleon
state~\cite{muta,bardeen},
\begin{eqnarray}
\label{ope13}
i \int d^4x e^{iqx}\langle p|
      T [ J_\mu(x) J_\nu(0) ] |p\rangle
= \nonumber \\
 \sum_n \frac{1}{x^n}
\left[ e_{\mu \nu}(C_{L,n}^q A^q_n +C_{L,n}^G A^G_n)
+d_{\mu \nu} (C_{2,n}^q A^q_n + C_{2,n}^G A^G_n)  \right]+ O(\frac{1}{Q^2})\ ,
\end{eqnarray}
where in the ${\overline {\rm MS}}$ scheme, the Wilson coefficients
are given in Table (\ref{tab:wilson}). For a proton state
normalized as $\langle p | p \rangle = (2\pi)^3 2 p_0 \delta^3(0)$,
we have
\begin{eqnarray}
\label{ope14}
A_n^q(\mu) & = & 2 \int_0^1 dx ~x^{n-1} [ q(x,\mu) +(-1)^n \bar{q}(x,\mu)]
\nonumber \\
A_n^G(\mu) & = & 2 \int_0^1 dx ~x^{n-1} G(x,\mu),
\end{eqnarray}
where $q(x,\mu)~(G(x,\mu)) $ is the quark (gluon) distribution function
inside the proton at the scale $\mu^2$.
Comparing eq.~(\ref{ope13}) multiplied by $\frac{\rho_n}{2m}$ with
the leading term in eq.~(\ref{ope10}) in the Bjorken limit, we
find,
\begin{eqnarray}
\label{ope16}
A^{2k+2,2}_{2k}
 & = & C_{2,2k}^q A^q_{2k} +  C_{2,2k}^G A_{2k}^G\ ,\nonumber \\
C^{2k+2,2}_{2k}
 & = & C_{L,2k}^q A^q_{2k} +  C_{L,2k}^G
 A_{2k}^G \ .
\end{eqnarray}

\subsection{Form of $A,B$ for $\tau=4$ in the linear density approximation}

The contributions due to $\tau=4$ operators have been computed
in~\cite{Jaffe} and those due to the mass dependent $\tau=4$
operator in~\cite{L94}
\begin{eqnarray}
\label{ope17}
i \int d^4x e^{iqx}\langle p|
      T [ J_\mu(x) J_\nu(0) ] |p\rangle \rightarrow  \nonumber \\
\frac{1}{x^2 Q^2} \left[ d_{\mu \nu} (H^1+ \frac{5}{8}H^2+ \frac{1}{16}H^g
-\frac{13}{4}H^m) + e_{\mu \nu} (\frac{1}{4}H^2 -\frac{3}{8}H^g -\frac{9}{2}
H^m ) \right],
\end{eqnarray}
where, for a current of $J^{Q_q}_\mu=\bar{q} Q_q\gamma_\mu q$, the
$H$'s  are the spin independent parts (such as the $A$ in
eq.~(\ref{ope9})) of the following operators,
\begin{eqnarray}
\label{ope18}
H^1_{\mu \nu} & = &\langle g^2 (\bar{q} \gamma_\mu \gamma_5
\frac{\lambda^a}{2} Q_q q)( \bar{q} \gamma_\nu \gamma_5
\frac{\lambda^a}{2} Q_q q) \rangle\ , \nonumber \\
H^2_{\mu \nu} & = &\langle g^2 (\bar{q} \gamma_\mu
\frac{\lambda^a}{2} Q_q^2 q)(\sum_{q=u,d,s} \bar{q} \gamma_\nu
\frac{\lambda^a}{2}  q) \rangle\ , \nonumber \\
H^g_{\mu \nu}& = &\langle ig (\bar{q} \{ D_\mu, ^*G_{\nu \alpha}
\}  \gamma^\alpha \gamma_5 Q_q^2 q \rangle \ ,\nonumber \\
H^m_{\mu \nu} & = & \langle \bar{q} D_\mu D_\nu m_q Q_q^2 q \rangle\ .
\end{eqnarray}
Comparing eq.~(\ref{ope17}) multiplied by $\frac{\rho_n}{2m}$ with
the $s=2, \tau=4$ part of eq.~(\ref{ope10}), we find,
\begin{eqnarray}
\label{ope19}
A^{6,4}_2 & = & H^1+\frac{5}{8} H^2 +\frac{1}{16} H^g - \frac{13}{4} H^m\ ,
\nonumber \\
C^{6,4}_2 & = & \frac{1}{4} H^2 -\frac{3}{8}H^g -\frac{9}{2} H^m\ .
\end{eqnarray}

The matrix elements of these operators are not all known. However,
the values of two different combinations in the $F_2$ and $F_L$
directions have been obtained by one of us~\cite{CHKL93} by
analyzing the  DIS data at CERN and SLAC.  In order to proceed, we
make the following assumption~\cite{CHKL93}:
\begin{eqnarray}
\label{ope20}
 { \langle \bar{d} \Gamma_\mu \Delta_\nu d \rangle \over
 \langle \bar{u} \Gamma_\mu \Delta_\nu u \rangle }
 =
{ \langle \bar{d} \gamma_\mu D_\nu d \rangle \over
 \langle \bar{u} \gamma_\mu D_\nu u \rangle }
\equiv \beta,
\end{eqnarray}
where $\Gamma_\mu$ is some gamma matrix and $\Delta_\nu$ is an
isospin singlet operator and we take $\beta=0.48$. Now, it is
possible to uniquely determine the matrix elements appearing in the
$\rho,\omega$ meson sum rules. To see this, let us rewrite the
scalar part of the matrix elements in eq.~(\ref{ope18}),
\begin{eqnarray}
H^1 & = & Q_u^2 K_u^1 +Q_d^2 K_d^1 -(Q_u -Q_d)^2 K_{ud}^1/2\ ,
\nonumber \\
H^2 & = & Q_u^2 K_u^2 +Q_d^2 K_d^2\ ,
\nonumber \\
H^g & = & Q_u^2 K_u^g +Q_d^2 K_d^g\ ,
\label{ope21}
\end{eqnarray}
where we consider only the $u,d$ sector and neglect the quark
masses $m_u ,m_d$. The $K's$ are the matrix elements obtained from
eq.~(\ref{ope18}),
\begin{eqnarray}
\label{ope22}
K_u^i & = & \frac{2}{m^2} \langle p|  {\bar u} \Gamma_+^i \Delta_+^i
 u | p \rangle, ~~i=1,2\ , \nonumber \\
K_u^g & = & \frac{2ig}{m^2} \langle p|  {\bar u} \{ D_+, {}^*
G_{+ \mu} \} \gamma^\mu \gamma_5
 u | p \rangle\ ,  \nonumber \\
K_{ud}^1 & = & \frac{2}{m^2} \langle p| 2(  {\bar u} \Gamma_+^1
 u) (  {\bar d} \Gamma_+^1  d)| p \rangle \ .
\end{eqnarray}
Here, $\Gamma_\mu^1=\gamma_\mu \gamma_5  \frac{\lambda^a}{2}, \Gamma_\mu^2
=\gamma_\mu \frac{\lambda^a}{2}, \Gamma_+ =(\Gamma_0 + \Gamma_3)/\sqrt{2}$
and $\Delta_\mu^i=\bar{u}\Gamma_\mu^i u +\bar{d}\Gamma_\mu^i d$ is a flavor
singlet operator.

Using the DIS data for $F_L,F_2$ on proton and neutron targets, one
can now determine the three combinations
\begin{eqnarray}
Z_1 & = & K_u^1+\frac{5}{8} K_u^2 + \frac{1}{16} K_u^g
\simeq -0.06~~ {\rm GeV}^2\ ,\nonumber \\
Z_2 & = & K_u^2-\frac{3}{2} K_u^g \simeq 0.56~~ {\rm GeV}^2\
,\nonumber \\ Z_3 & = & K_{ud}^1/2 \simeq -0.04 ~~ {\rm GeV}^2\ .
\label{ope23}
\end{eqnarray}
The operator product expansion for $\Pi_{L,T}$ is then given by
\begin{eqnarray}
\label{ope24}
\Pi^1_{L} & = &  \frac{m \cdot \rho_n}{\omega^8}
  \biggl[ \frac{1}{4}(1+\beta) \left( \frac{1}{2} Z_2  \right)\biggl]\ ,
 \nonumber  \\
\Pi^1_{T} & = &  \frac{m \cdot \rho_n}{\omega^8}
  \biggl[\frac{1}{4}(1+\beta) \left( -2 Z_1 +\frac{1}{2} Z_2  \right) +
\frac{1 \pm 1}{2} Z_3 \biggl].
\end{eqnarray}
In the last term in $\Pi_{L,T}$, the $+$ sign refers to the $\rho$
and the $-$ sign to the $\omega $ meson.

\subsection{Final form of the OPE}

The final form of the OPE for $\Pi^1$ can now be obtained by using
eqs. (\ref{ope11}, \ref{ope12}, \ref{ope16}) and (\ref{ope24}):
\begin{eqnarray}
\Pi_{L,T}^1(\omega)/\rho_n= {b^{L,T}_2 \over \omega^6} + {b^{L,T}_3
\over \omega^8}\ .
\label{ope25}
\end{eqnarray}

For the longitudinal (L) and transverse (T) parts one
finds~\cite{Lee98c,LK98},
\begin{eqnarray}
b_2^T & =& ( \frac{1}{2}C_{2,2}^q -\frac{1}{2}C_{L,2}^q  )
mA_2^{u+d}  +
(C_{2,2}^G -C_{L,2}^G) mA_2^{G}\ , \nonumber \\
b_3^T & = &  ( \frac{9}{4}C_{2,4}^q -\frac{5}{2}C_{L,4}^q  )
m^3A_4^{u+d}  +
(\frac{9}{2}C_{2,4}^G -5C_{L,4}^G) m^3A_4^{G} \nonumber \\
& & + \frac{1}{2} m \left[
(1+\beta)(-Z_1+\frac 14 Z_2) +(1\pm1)Z_3
\right]\ , \nonumber \\
b_2^L & =&  -\frac{1}{2}C_{L,2}^q mA_2^{u+d}  -C_{L,2}^G mA_2^{G}, \nonumber
\\
b_3^L & = & (\frac{1}{2}C_{2,4}^q -\frac{5}{2}C_{L,4}^q) m^3A_4^{u+d}  +
(C_{2,4}^G -5C_{L,4}^G) m^3A_4^{G} + \frac{m}{8}(1+\beta)Z_2,
\label{ope26}
\end{eqnarray}
where again the $\pm$ refers to  the $\rho$ and $\omega$ mesons,
respectively.

\subsection{Infinite sum of twist-2 contribution}

Since the matrix elements of all the $\tau=2$ operators are known,
one can sum the higher order $\frac{1}{\omega^n}$( $n >6$) twist-2
correction. Here we present the result to lowest order in
$\alpha_s$, i.e., including only the quark part of
eq.~(\ref{ope16}). By substituting this into eq.~(\ref{ope10}) and
evaluating the sums we find,
\begin{eqnarray}
\label{ope27}
\Pi^1_L & = & \frac{\rho_nm}{2\omega^6} \int_0^1dx x
 {(1+\frac{m^2 x^2}{\omega^2} )^2  \over (1-\frac{m^2 x^2}{\omega^2} )^2 }
[ q(x)+\bar{q}(x)] \nonumber \\
 & & -
\frac{\rho_nm}{2\omega^6} \int_0^1 dx \left[
8 {(1+\frac{m^2 x^2}{\omega^2} )^6 \over (1-\frac{m^2 x^2}{\omega^2} )^6 }
-12 {(1+\frac{m^2 x^2}{\omega^2} )^4 \over (1-\frac{m^2 x^2}{\omega^2} )^4 }
+3  {(1+\frac{m^2 x^2}{\omega^2} )^2 \over (1-\frac{m^2 x^2}{\omega^2} )^2 }
\right] \nonumber \\ && \hspace*{5cm} \times
 \int_1^x dy [ q(y)+\bar{q}(y)]
\nonumber \\
& & +
\frac{\rho_nm}{2\omega^6} \int_0^1 dx x \left[
8 {(1+\frac{m^2 x^2}{\omega^2} )^6 \over (1-\frac{m^2 x^2}{\omega^2} )^6 }
-16 {(1+\frac{m^2 x^2}{\omega^2} )^4 \over (1-\frac{m^2 x^2}{\omega^2} )^4 }
+8  {(1+\frac{m^2 x^2}{\omega^2} )^2 \over (1-\frac{m^2 x^2}{\omega^2} )^2 }
\right] \nonumber \\ && \hspace*{5cm} \times
 \int_1^x dyy^{-1}  [ q(y)+\bar{q}(y)]\ \nonumber, \\[15pt]
\Pi^1_T & = & \frac{\rho_nm}{\omega^6} \int_0^1 dx x
{ (1+\frac{m^2 x^2}{\omega^2} ) \over (1-\frac{m^2 x^2}{\omega^2} )^2 }
[ q(x)+\bar{q}(x)]  \nonumber \\
 & & -
\frac{\rho_nm}{\omega^6} \int_0^1 dx \left[
16 {(1+\frac{m^2 x^2}{\omega^2} )^5 \over (1-\frac{m^2 x^2}{\omega^2} )^6 }
-14 {(1+\frac{m^2 x^2}{\omega^2} )^3 \over (1-\frac{m^2 x^2}{\omega^2} )^4 }
+  {(1+\frac{m^2 x^2}{\omega^2} ) \over (1-\frac{m^2 x^2}{\omega^2} )^2 }
\right] \nonumber \\ & & \hspace*{5cm} \times
 \int_1^x dy  [ q(y)+\bar{q}(y)]
\nonumber \\
& & +
\frac{\rho_nm}{\omega^6} \int_0^1 dx x  \left[
16 {(1+\frac{m^2 x^2}{\omega^2} )^5 \over (1-\frac{m^2 x^2}{\omega^2} )^6 }
-20 {(1+\frac{m^2 x^2}{\omega^2} )^3 \over (1-\frac{m^2 x^2}{\omega^2} )^4 }
+4  {(1+\frac{m^2 x^2}{\omega^2} ) \over (1-\frac{m^2 x^2}{\omega^2} )^2 }
\right] \nonumber \\ & &
 \hspace*{5cm} \times \int_1^x dyy^{-1}  [ q(y)+\bar{q}(y)]\ .
\end{eqnarray}
In the following section we use this partial summation to estimate
the effect of higher dimensional ($n > 6$) operators on the
constraint.

\section{Constraints}

We now make a Borel transformation of $\omega^2$ times
eq.~(\ref{ope25}) and examine the dispersion relation,
\begin{eqnarray}
\rho_n \left(
 \frac{b_2}{M^2} -\frac{b_3}{2M^4}  \right)
= \int d\omega^2~ \rho^1(\omega^2 )~ \omega^2~ e^{-\omega^2 /M^2} -\rho_n b_{scatt}\ .
\label{ope28}
\end{eqnarray}
The reason for multiplying by $\omega^2$ is to get rid of any
possible subtraction constants proportional to $1/\omega^2$.  There
could be further subtraction constants proportional to $\rho_n
b_{scatt}/\omega^4$, which we include. In a given model
calculation, the corresponding value of $b_{scatt}$ should be
computed. The contribution from particle-hole intermediate states
is $b_{scatt}=1/4m $ for the longitudinal $\rho,\omega$ meson and
zero for the transverse parts. (This is a non-trivial contribution
in the sense discussed above.) After the Borel transformation, the
contribution from dimension=$n$ operators to the OPE is divided by
$(n-2)!$. This improves the convergence of the expansion, and
enhances the contribution of the low-mass region.

In Fig.~1, we show the OPE side of eq.~(\ref{ope28}) as a function
of $M^2$ for the $\rho$ meson. The solid lines shows the full
calculation including dimension 6 operators, while the dot-dashed
line shows the OPE without the twist-4 contributions and the long
dashed line without dimension 6 contributions.
At $M_{min}^2 \sim 0.8~ {\rm GeV}^2$ ($M_{min}^2 \sim 2.5~{\rm
GeV}^2$), the contributions of dimension 6 operators is
approximately 40\% of those from dimension 4 operators for
transverse (longitudinal) polarization.  Therefore, the OPE should
be reliable only for values of the Borel mass above these, where
one expects the contributions from higher dimensional operators to
be suppressed.  We also show (short-dashed line) the OPE including
the contributions from the infinite sum of twist-2 operators,
eq.~(\ref{ope27}). The corresponding Borel transform is given in
Appendix B. The fact that the contribution of twist 2 operators of
dimension larger than 6 is very small for $M^2> M^2_{min}$,
indicates that the truncated OPE is indeed reliable above the
minimum Borel mass.

Thus, the constraints on the spectral density are given by
eq.~(\ref{ope28}), for $M^2> M^2_{min}$. The Borel transformation
also introduces an exponential weighting factor of the spectral
density $\exp(-s/M^2)$, which for small values of Borel mass,
suppresses the contributions at large energy. This has the
advantage that in practical calculation, one can concentrate on the
change of the spectral density near and below the vector meson mass
region, while the contribution from higher energies can be
effectively taken into account by a continuum. This approximation
will be valid for Borel masses below some maximum value $M_{max}$.
We estimate $M_{max}$ by requiring that the continuum contribution
should be less than that of the low energy region.

\section{Comparison With Hadronic Model Calculations}

A possible non-trivial momentum dependence of the vector-meson self
energy in nuclear matter has recently been extensively discussed.
Such a momentum dependence arises naturally through the energy
dependence of the $\rho$-nucleon scattering amplitude. In this
context $\rho$-nucleon interactions in p-wave channels play an
important role~\cite{FP97,RCW97,PPLLM98}. In a complementary
approach, the momentum dependence due to in-medium modifications of
the $\rho$-meson pion cloud has been explored~\cite{UBRW98}. We
note that in none of these models is the momentum dependence
directly constrained by data. Therefore it is useful to confront
them with the constraints from QCD sum rules. Here we explore
models of the type studied in refs.~\cite{FP97,PPLLM98}, where the
$\rho$-nucleon interaction is mediated by s-channel resonance
exchanges.

To make a systematic comparison with our constraints, we have to
first relate the correlation operator eq.~(\ref{ope1}) and the
polarization functions eq.~(\ref{ope2}) to the $\rho$ meson
self-energy $\Sigma_{L,T}(\omega,{\bf q})$, as computed in a
hadronic model. To this end, we invoke vector-meson-dominance
~\cite{VMD}
\begin{eqnarray}
\label{com1}
\Pi_{L,T}(\omega,{\bf q}) = - \frac{1}{q^2} \frac{m_\rho^4}{g_\rho^2}
 { 1 \over q^2 -m_\rho^2 + i\Gamma m_\rho - \Sigma_{L,T} (\omega, {\bf q}) }\ ,
\end{eqnarray}
where $g_\rho=6.05$ is the $\rho\pi\pi$ coupling constant.

The next step is to extract the lowest order nontrivial ${\bf q}^2$
dependence to linear order in density.  To this order, the self
energy is of the following generic form for p-wave
resonances~\cite{FP97,PPLLM98},
\begin{eqnarray}
\label{com2}
\Sigma_T= {\bf q}^2 \left( \frac{f_{RN\rho}}{m_\rho} \right)^2
 [F( {\bf q})]^2 \frac{S_\Sigma}{2} \rho_n {E_R( {\bf q})-m_N
 \over\omega^2 - (E_R ({\bf q}) -m_N)^2} \ .
\end{eqnarray}
In this section we consider only transversely polarized vector
mesons; in the longitudinal channel the expected momentum
dependence is much weaker~\cite{Lee98c,PPLLM98}. For s-wave
resonances~\cite{PPLLM98} the leading ${\bf q}^2$ dependence in
(\ref{com2}) is replaced by a factor proportional to $\omega^2$,
\begin{eqnarray}
\label{com3}
\Sigma_T= \omega^2 \left( \frac{f_{RN\rho}}{m_\rho} \right)^2
 [F( {\bf q})]^2 \frac{S_\Sigma}{2} \rho_n {E_R( {\bf q})-m_N \over
 \omega^2 - (E_R ({\bf q}) -m_N)^2 }\ ,
\end{eqnarray}
where the form factor is parametrized as
\begin{eqnarray}
\label{com4}
 F( {\bf q})= { \Lambda^2 \over \Lambda^2 +{\bf q}^2 }  ~~~
{\rm with} ~~ \Lambda= 1.5~ {\rm GeV}\ ,
\end{eqnarray}
and
\begin{eqnarray}
\label{com5}
E_R ({\bf q})= \sqrt{ m_R^2 + {\bf q}^2 } - \frac{i}{2} \Gamma_R\ .
\end{eqnarray}
Here $m_N, m_R$ are the masses of the nucleon and the resonances,
and $\Gamma_R$ is the total width of the resonance and its momentum
dependence will be modeled as in ref.~\cite{FP97},
\begin{eqnarray}
\label{com6}
\Gamma_R(\omega, {\bf q})=\Gamma_R^0 \left(
 {p \over p_R} \right)^{2l+1} \ .
\end{eqnarray}
In the phase-space factor, $l=0 (1)$ for s- (p-)wave resonances and
\begin{eqnarray}
p= \sqrt{ [ (s-m_N^2-m_\pi^2)^2-4 m_N^2 m_\pi^2]/4s } \ , \nonumber \\
p_R= \sqrt{ [ (m_R^2-m_N^2-m_\pi^2)^2-4 m_N^2 m_\pi^2]/4m_R^2 }\ ,
\end{eqnarray}
where $ s=(\omega+m_N)^2-{\bf q}^2$. However, in the present
context the effect of the momentum dependence of the width turns
out to be weak.

As can be seen from eq.~(\ref{com2}), the contribution from p-wave
resonances is proportional to ${\bf q}^2$. Hence its contribution
to Im$\Pi^1$ is  trivially obtained by neglecting the remaining
${\bf q}^2$ dependence.   In contrast, eq.~(\ref{com3}) is
proportional to $\omega^2$.  The contribution of such terms to
Im$\Pi^1$ is most easily obtained by replacing $\omega^2$ by  $q^2
+ {\bf q}^2$ and expanding to leading order in ${\bf q}^2$ keeping
$q^2$ fixed. Finally, we obtain
\begin{eqnarray}
\label{com7}
{\rm Im} \Pi^1 (\omega^2)_{res}&  = &-\frac{1}{\omega^2}
\frac{m_\rho^4}{g_\rho^2}
 {\rm Im} \Bigg [{ 1 \over (\omega^2 -m_\rho^2 + i m_\rho
 \Gamma_\rho )^2}
\times
\nonumber \\
 & & \Biggl \{ \sum_{s,p-wave}  \left( \frac{f_{RN\rho}}{m_\rho} \right)^2
 \frac{S_\Sigma}{2} \rho_n
 {m_R-m_N -\frac{i}{2} \Gamma_R \over \omega^2 - (m_R -m_N -\frac{i}{2}
\Gamma_R)^2 }
\nonumber \\
& & + \sum_{s-wave} \left( \frac{f_{RN\rho}}{m_\rho} \right)^2
 \frac{S_\Sigma}{2} \rho_n  \omega^2
\Bigg (
 {- \frac{1}{\Lambda^2}(  m_R-m_N -\frac{i}{2} \Gamma_R)
 \over \omega^2 - (m_R -m_N -\frac{i}{2} \Gamma_R)^2 } \nonumber \\
& &
+ { \frac{1}{2m_R} \over \omega^2 - (m_R -m_N -\frac{i}{2}
\Gamma_R)^2 }
+ { - \frac{1}{m_R}(  m_N +\frac{i}{2} \Gamma_R)(m_R-m_N -\frac{i}{2} \Gamma_R)
 \over ( \omega^2 - (m_R -m_N -\frac{i}{2}
\Gamma_R)^2)^2 } \Bigg ) \Biggl \}\Bigg ] \ .
\end{eqnarray}
where we have neglected the small contribution due to the momentum
dependence of $\Gamma_R$.

The contribution from the higher energy region is modeled using the
following assumption. In the vacuum QCD sum rule approaches, the
spectral density is approximated by a pole and a continuum.  Since
we make a hadronic calculation of the changes in the pole, we will
add the simplest change in the continuum and allow a non-trivial
change in the continuum threshold,
\begin{eqnarray}
c_0 \theta \left( q^2-(s_0+s' {\bf q}^2) \right) ,
\end{eqnarray}
where $s'$ is a parameter to be determined.  The parameter $c_0$,
which characterizes the strength of the high-energy continuum, is
determined by QCD duality. This excludes a simple ${\bf q}^2$
dependence, such as $c_0\rightarrow c_0 + c'{\bf q}^2$. We thus
find the following contribution to the spectral density,
\begin{eqnarray}
\label{com8}
{\rm Im} \Pi^1 (\omega^2)_{cont}=-c_0 s' \delta(\omega^2-s_0)
\end{eqnarray}
The phenomenological spectral density $\rho^1$ is then given by the
sum of eq.~(\ref{com7}) and eq.~(\ref{com8}), i.e., $\rho^1 =
(1/\pi) ({\rm Im}\Pi^1_{res}+{\rm Im}\Pi^1_{cont})$. The parameter
$s'$ in eq.~(\ref{com8}) is determined by minimizing the difference
between the left and right hand side in eq.~(\ref{ope28}). We note
that the parameter $s_0 = 1.43$ GeV$^2$ is fixed by the QCD sum
rule~\cite{HL92} at ${\bf q} = 0$.

\subsection{A `standard' set of model parameters}

We first explore the model employed by Peters {\it et
al.}~\cite{PPLLM98}. The corresponding parameters are given in
Table 2. We choose this set, since it is representative for the
sets used also by other groups, with similar values for the
parameters (see e.g. refs.~\cite{RCW97,RUBW97,Klingl}).
In Fig.~2 we compare the OPE with the phenomenological side, i.e.,
the left and the right hand sides of eq.~(\ref{ope28}), including
both the resonance and continuum contributions in the
phenomenological spectral density. The best fit was obtained in the
Borel window $0.5 - 2.0$ GeV${}^2$, by adjusting the parameter $s'$
of the continuum. We find a huge discrepancy between the OPE and
the phenomenological side. The major contributions to the
phenomenological side are due to the nucleon as well as the
$\Delta(1232)$, $N(1520)$, $N(1680)$, $N(1720)$ and $\Delta(1905)$
resonances. For comparison we show the right hand side for only
these six states (denoted as `selected subset' in Fig.~2) as well
as some of the individual contributions. Clearly, the very large
effect of the $\Delta(1232)$ is mainly responsible for the
disagreement. In the best fit, its contribution, which in itself is
much larger than the OPE side, cannot be compensated by the
continuum over a reasonable range of Borel masses. The fact that
the continuum lies at a much higher energy than the $\Delta(1232)$,
leads to a characteristic difference in the dependence on the Borel
mass. Thus, in most of the Borel window, the best fit lies far
outside the estimated boundary of uncertainty shown in Fig.~1.

In Fig.~3, we show the corresponding spectral density $\rho^1$. The
contribution of a particular resonance is identified by extracting
the corresponding term in the curly brackets in eq.~(\ref{com7}).
Also here the dominance of the $\Delta(1232)$ is apparent.
Moreover, in the constraint equation (\ref{ope28}), the
contribution of the $\Delta(1232)$ is further enhanced compared to
the other structures at higher energy due to the exponential
factor. Thus, we conclude that the strong effective $\Delta  N
\rho$ coupling originally employed by Rapp {\em et al.}~\cite{RCW97}
is incompatible with the constraint from QCD sum rules.

\subsection{A consistent model}

As noted above, the most important contributions to the sumrule are
due to the selected set of resonances. Consequently, in the
following analysis, we will include only these baryon states. Let
us first recall how the coupling constants $f_{RN\rho}$ are fixed.
The couplings for the $N(1720)$ and the $\Delta(1905)$ are
determined from their decays into a nucleon and a $\rho$
meson~\cite{FP97}. Hence, these couplings should be reliable for
the process $\rho+ N \rightarrow resonance$ considered here. If the
mass of the resonance lies below the $\rho$ N threshold, however,
the corresponding coupling constant can clearly not be determined
in this way. Consequently, in many models the $\rho$ N couplings to
the nucleon and the $\Delta(1232)$, are obtained e.g. from the Bonn
potential~\cite{Bonn}. An important point, which apparently has
been overlooked, is that in the Bonn potential the coupling
constant $f_{RN\rho}$ is multiplied by a form factor~\cite{Bonn},
\begin{eqnarray}
F_\rho({\bf q}^2)= \biggl( { \Lambda_\rho^2-m_\rho^2 \over
\Lambda_\rho^2+ {\bf q}^2 } \biggl)^{n_\rho}.
\label{ff}
\end{eqnarray}
where $\Lambda_\rho=1.4$GeV and $n_\rho=1$ for the $NN \rho$ and
$n_\rho=2$ for the $\Delta N\rho$ vertex. Hence,  the effective
value of the coupling constant at ${\bf q}=0$ is reduced by 0.7 for
the nucleon and by 0.49 for the $\Delta(1232)$. Since the self
energy is proportional to the square of the coupling constant and
the corresponding form factor, the contribution of the nucleon
($\Delta(1232)$) is reduced by a multiplicative factor of 0.49
(0.24). This reduction of the effective coupling constant has been
left out in previous calculations.

The resonances $N(1520)$ and $N(1680)$ are situated below the
nominal threshold for the decay into a nucleon and a hypothetical
zero-width $\rho$ meson. Thus, they can decay into a nucleon and a
$\rho$ meson only because of the finite widths of the particles
involved. Since in this case only the low-mass tail of the $\rho$
meson is seen in the two-pion spectrum~\cite{manley}, a reliable
identification of the $\rho$ is very difficult. Consequently, the
determination of the corresponding coupling constants is rather
uncertain. In order to explore possible consequences of this
uncertainty, we also show results for the case where the two
coupling constants vanish.

Let us now confront this set of coupling constants with the QCD sum
rules. For the nucleon and the $\Delta(1232)$, we use the couplings
of the Bonn potential, with the corresponding form factors,
extrapolated into the time-like region
\begin{eqnarray}
\label{fc}
F^{\,c}_\rho({q}^2)= \left\{ \begin{array}{ll}
 \biggl( { \Lambda_\rho^2-m_\rho^2 \over
\Lambda_\rho^2- q^2 } \biggl)^{n_\rho} & \mbox{ if $q^2< m_\rho^2$}
\\
1+ \biggl({q^2-m_\rho^2 \over \Lambda_\rho^2} \biggl)^{n_i}
 & \mbox{ otherwise. }
\end{array}
\right.
\end{eqnarray}
Thus, for $q^2< m_\rho^2$ we use the covariant generalization of
(\ref{ff}). However, close to the pole at $q^2 = \Lambda_\rho^2$
this form clearly makes no sense. For $q^2> m_\rho^2$ we therefore
employ the simple form shown in the second line of (\ref{fc}), with
the exponent $n_i
= 0, 1$ or $2$. Fortunately, due to the
exponential weighting, the QCD sum rule (\ref{ope28}) is
insensitive to the precise form of $F^{\,c}$ in this range of
energies; the real parts of $\Pi^1$ obtained from the right-hand
side of (\ref{ope28}) for different values of $n_i$ are almost
indistinguishable. Thus, for the nucleon and the $\Delta(1232)$ we
replace the coupling constant $f_{RN\rho}$ in eq.~(\ref{com7}) by
$f^*_{RN\rho}(\omega^2)=F^{\,c}_\rho(\omega^2)f_{RN\rho}$.

For the $\Delta(1905)$ and the $N(1720)$ we use the coupling constants
$f_{RN\rho}$ determined in ref.~\cite{FP97}. Since the corresponding
form factors are normalized to unity for zero three-momentum, they do
not affect the leading ${\bf q}^2$ dependence. Furthermore, a possible
frequency dependence can also be neglected, because, in contrast to
the $\Delta(1232)$, these resonances contribute significantly only in
the kinematical region where the coupling constants are determined,
i.e., for almost on-shell $\rho$ mesons. The coupling constants used
are given in Table III. Note that here we use the $NN\rho$ and
$\Delta(1232) N \rho$ coupling constants of ref.~\cite{Bonn}, which
differ slightly from those used in ref.~\cite{PPLLM98}.

In Fig.~4 we show the effect of a consistent treatment of the
$\Delta N\rho$ form factor on the spectral density. As expected,
the contribution due to the $\Delta(1232)$ is substantially
reduced. This reduction is also reflected in an improved agreement
with the constraint equation, as shown in Fig.~5. There, the
distance between the two solid lines, which correspond to the OPE
with and without the dimension 6 operators, represent the
uncertainty of the left-hand side of (\ref{ope28}). The long-dashed
line is the right-hand side without the form factor for the delta
and the nucleon, while the dashed line shows the result including
form factors of the form eq.~(\ref{fc}) with $n_i=0$. (Since, as
noted above, the dependence on $n_i$ is minimal, we do not show
results for $n_i=1,2$.) Furthermore, the dotted line shows the
result of including the form factors but with vanishing couplings
to the $N(1520)$ and $N(1680)$ resonances.  Thus, the range between
the dashed and dotted lines indicate the uncertainty due to the
poorly known $N(1520)$ and $N(1680)$ coupling constants.

Using the criteria discussed in Sect.~III, we obtain a Borel
window, where the constraint equation can be applied, of $0.8~{\rm
GeV}^2 < M^2 < 1.8~{\rm GeV}^2$. Clearly, there is a good agreement
between the OPE and the phenomenological calculation, when the
$\Delta N\rho$ form factor is consistently taken into account. We
note that most of the disagreement is removed already with a static
(frequency independent) form factor, i.e., by renormalizing the
effective coupling strength by 0.7 for the nucleon and by 0.49 for
the $\Delta$. The resulting phenomenological side is shown by the
dash-dotted line in Fig.~5. However, the frequency dependence, as
parameterized in (\ref{fc}), leads to a further improvement (see
Fig.~5).

Our work indicates that the $\Delta N \rho$ coupling used in many
model calculations is unphysical. The Bonn potential offers no
justification for the very strong effective couplings used.
Furthermore, such couplings violate the constraints obtained from
QCD sum rules. Thus, it is necessary to re-analyze those
calculations of the low-mass dilepton spectrum in heavy-ion
collisions, where the $\Delta N\rho$ coupling was treated
inconsistently.

\section{Discussion}

We have presented a model independent constraint for the three
momentum dependence of the vector meson spectral function in a
nuclear medium. We have shown that the momentum dependence of the
spectral density obtained in model
calculations~\cite{RCW97,RUBW97,PPLLM98,Klingl} is too strong to
satisfy the constraint derived from QCD sum rules. The main source
of the discrepancy is the large contribution from the
$\Delta(1232)$. In these models, the coupling of the rho meson to
the nucleon and the $\Delta(1232)$ is not accounted for in a
consistent manner. While the coupling constant is taken e.g. from
the Bonn potential, the form factor is not; the normalization of
the form factor differs considerably from that, which is used in
the Bonn potential. This leads to an effective coupling strength
which is approximately a factor two too large. When the original
Bonn form factor is employed, the contribution of the
$\Delta(1232)$ is substantially reduced and we find good agreement
with the constraint equation.

In view of this it would be important to reexamine the hadronic
model calculations that are used to interpret the low-mass dilepton
enhancement found in relativistic heavy-ion collisions, taking the
form factors properly into account. The unphysically large coupling
to the $\Delta(1232)$ gives rise to a spurious enhancement of the
vector meson spectral density for low invariant masses.

\section*{Acknowledgments}

We would like to thank T. Hatsuda, F. Klingl,  M. Lutz, W. Weise
for useful discussions. The work of S. H. Lee is supported by KOSEF
through grant no. 971-0204-017-2 and 976-0200-002-2,  by the Korean
Ministry of Education through grant no. 98-015-D00061 and by a
Humboldt fellowship. The work of  H. Kim is supported by a Research
Fellowships of the Japan Society for the Promotion of Science.

\begin{appendix}

\section{Coefficients of trace terms}

Here we specify the trace terms in detail. As discussed in the
text,
\begin{eqnarray}
A_{\mu_1 \cdot \cdot \cdot \mu_{2k}}= \left(
 \sum_{j=0}^k a_j p^{2j} [g_{\mu_1 \mu_2}
\cdot \cdot \cdot g_{\mu_{2j-1} \mu_{2j}}
   p_{\mu_{2j+1}}
\cdot \cdot \cdot p_{\mu_{2k}} + permutations
 ] \right) A_{2k}\ ,
\end{eqnarray}
where for a given value of $k$ and $j$ there are~\cite{GP76}
$\frac{(2k)!}{(2k-2j)!2^j j!}$ different terms and $a_j=(-1)^j
\frac{(2k-j)!}{2^j (2k)!}$.

Now consider the two indices $\mu_1,\mu_2$. They can enter either
as indices of the metric tensor $g_{\mu\nu}$ or of the nucleon four
momentum $p_\mu$. Depending on how the two indices appear, we can
distinguish four types of terms
\begin{eqnarray}
g_{\mu_1 \mu_2} & \times & \left[ (j-1)~ {\rm factors
 }~ (g_{\mu \nu}) \right] \times \left[ 2(k-j) ~ {\rm
 factors }~ (p_{\mu})
\right] \nonumber \\
g_{\mu_1 \kappa}g_{\mu_2 \lambda} & \times & \left[ (j-2)~ {\rm factors
 }~ (g_{\mu \nu}) \right] \times \left[ 2(k-j) ~ {\rm factors }~
 (p_{\mu})\right] \nonumber \\
g_{\mu_1 \kappa}p_{\mu_2} & \times & \left[ (j-1)~ {\rm factors
 }~ (g_{\mu \nu})\right]  \times \left[ (2k-2j-1)
 ~ {\rm factors }~ (p_{\mu}) \right]
 \nonumber \\
p_{\mu_1}p_{\mu_2} & \times & \left[ (j)~ {\rm factors
 }~ (g_{\mu \nu})\right]  \times  \left[ (2k-2j-2)
 ~ {\rm factors}~ (p_{\mu}) \right] .
\end{eqnarray}
There are ${(2k-2)! \over (2k-2j)! 2^{j-1} (j-1)! } $ number of
terms of the first type, where the remaining indices are ordered
differently. Of the second type there are ${(2k-2)!
\over (2k-2j)! 2^{j-2} (j-2)! }$, of the third ${2(2k-2)!
\over (2k-2j-1)! 2^{j-1} (j-1)! }$ and of the fourth there are $
{(2k-2)!\over (2k-2j-2)! 2^{j} (j)! }$ number of terms. Using these
results, the infinite series of twist-2 contributions can be
summed.

\section{Borel Transformation of the infinite Sum}

The Borel transformations of eq.~(25) can be obtained from the
basic formula
\begin{eqnarray}
{\rm B.T.} \left( \frac{1}{(\omega^2-A)^s} \right)=(-1)^s
\frac{1}{(s-1)!} \frac{1}{(M^2)^{s-1}} e^{-A/M^2}
\end{eqnarray}
We find,
\begin{eqnarray}
{ {\rm B.T.} (\omega^2 \Pi^0) \over \rho_nm}   & = & \int_0^1 dx x
 \left[ \frac{1}{m^2 x^2}(1-e^{-m^2 x^2/M^2})+\frac{2}{M^2}e^{-m^2 x^2/M^2}
  \right]
[ q(x)+\bar{q}(x)]  \nonumber \\
 &+&
  \int_0^1 dx
 \Bigg [ -\frac{1}{m^2x^2} (1-e^{-m^2 x^2 /M^2}) +e^{-m^2 x^2/M^2}
  ( \frac{6}{M^2}- 12 \frac{m^2 x^2}{M^4} \nonumber \\
& &~~~ + \frac{8}{3} \frac{m^4 x^4}{M^6})
  \Bigg ]
\int_1^x dy [ q(y)+\bar{q}(y)]
\nonumber \\
&-&
  \int_0^1dx x
 e^{-m^2 x^2/M^2} \left ( \frac{8}{M^2}- 12 \frac{m^2 x^2}{M^4} + \frac{8}{3}
 \frac{m^4 x^4}{M^6} \right )
\int_1^x dyy^{-1} [ q(y)+\bar{q}(y)]\ ,
\\[18pt]
{ {\rm B.T.} (\omega^2 \Pi^1_L) \over \rho_nm}    & = &  \int_0^1dx x
 \left[ (\frac{1}{2M^2} -\frac{2}{m^2 x^2} ) +
 e^{-m^2 x^2/M^2} ( \frac{2}{m^2 x^2}+ \frac{2}{M^2})  \right]
[ q(x)+\bar{q}(x)] \nonumber \\
 & & +
 \int_0^1 dx \Bigg [
\frac{6}{m^2 x^2}( 1-e^{- m^2 x^2/M^2}) +
e^{-m^2 x^2/M^2} (-\frac{6}{M^2}-16 \frac{m^4 x^4}{M^6}\nonumber \\
& & ~~~~
+\frac{32}{3} \frac{m^6 x^6}{M^8} -\frac{32}{15} \frac{m^8 x^8}{M^{10}})
 +\frac{1}{2M^2} \Bigg ]  \int_1^x dy [ q(y)+\bar{q}(y)]
  \nonumber \\
& & +
 \int_0^1 dx x
e^{-m^2 x^2/M^2} \left ( \frac{32}{3} \frac{m^4 x^4}{M^6}
-\frac{32}{3} \frac{m^6x^6}{M^8} + \frac{32}{15} \frac{m^8 x^8}{M^{10}}
\right)  \nonumber \\ & & \hspace*{5cm}
  \times \int_1^x dyy^{-1}  [ q(y)+\bar{q}(y)]\ ,
 \\[18pt]
{ {\rm B.T.} (\omega^2 \Pi^1_T) \over  \rho_nm} & = & \int_0^1 dx x
 \left[ -\frac{1}{m^2 x^2}  + e^{-m^2 x^2/M^2}(\frac{1}{m^2 x^2}
 + \frac{2}{M^2} ) \right]
[ q(x)+\bar{q}(x)]  \nonumber \\
 & & +
 \int_0^1 dx \Bigg [
\frac{3}{m^2 x^2} (1-e^{-m^2 x^2/M^2})
+ e^{-m^2 x^2 /M^2}( -\frac{6}{M^2} + 36 \frac{m^2x^2}{M^4}\nonumber \\
& & ~~~-\frac{200}{3} \frac{m^4 x^4}{M^6}
 + 32 \frac{m^6 x^6}{M^8}
-\frac{64}{15} \frac{m^8 x^8}{M^{10}}) \Bigg ]  \int_1^x dy [ q(y)+\bar{q}(y)]
\nonumber \\
& & +
  \int_0^1 dx x  e^{-m^2 x^2/M^2}
\left ( -24\frac{m^2 x^2}{M^4} + \frac{176}{3} \frac{m^4 x^4}{M^6}
 -32 \frac{m^6 x^6}{M^8} + \frac{64}{15} \frac{m^8 x^8}{M^{10}}
\right ) \nonumber \\
& & \hspace*{5cm} \times
 \int_1^x dyy^{-1}  [ q(y)+\bar{q}(y)] \ .
\end{eqnarray}

\end{appendix}

\begin{table}[hbt]
\caption{Wilson coefficients for $\tau=2$.
$T(R)=\frac{f}{2},C_2(R)=\frac{4}{3}$.
\label{tab:wilson}}
\vspace*{0.5cm}
\begin{tabular}{l  l l}
 Wilson coefficient  & $n=2$ & $ n=4$
\\
\hline
 $ C_{2,n}^q = 1+ \frac{\alpha_s}{4\pi} B_{2,n}^{NS} $   &
 $ 1+ \frac{\alpha_s}{4\pi}( 0.44) $           &
 $ 1+ \frac{\alpha_s}{4\pi}(6.07) $
\\
 $ C_{L,n}^q = \frac{\alpha_s}{4 \pi} C_2(R) \frac{4}{n+1} $  &
 $ \frac{\alpha_s}{4\pi}\frac{16}{9}  $  &
 $ \frac{\alpha_s}{4\pi}\frac{16}{15} $
\\
 $ C_{2,n}^G = \frac{\alpha_s}{4 \pi} T(R) \frac{4}{f}
  [ \frac{4}{(n+1)} - \frac{4}{(n+2)} + \frac{1}{n^2}
  - \frac{n^2+n+2}{n(n+1)(n+2)}(1+\sum_{j=1}^n \frac{1}{j} )] $  &
 $ \frac{\alpha_s}{4\pi}(-\frac{1}{2})  $ &
 $ \frac{\alpha_s}{4\pi}(-\frac{133}{180}) $
\\
 $ C_{L,n}^G =
  \frac{\alpha_s}{4 \pi} T(R) \frac{1}{f} \frac{16}{(n+1)(n+2)} $  &
 $ \frac{\alpha_s}{4\pi} \frac{2}{3}  $  &
 $ \frac{\alpha_s}{4\pi} \frac{4}{15} $
\end{tabular}
\end{table}

\begin{table}
\caption{The parameters used in ref.~[31].  }

\begin{center}
\begin{tabular}{cccccc}
 & $I(J^P)$ & $\Gamma_{Total}$(GeV) & $l_\rho$ & $f_{RN\rho}$ & $S_\Sigma$  \\
\hline\hline
N(940)  & $\frac{1}{2}( \frac{1}{2}^+)$ & 0 & 1 & 7.7 & 4 \\
N(1520)  & $\frac{1}{2}( \frac{3}{2}^-)$ & 120 & 0&  7 & 8/3 \\
N(1650)  & $\frac{1}{2}( \frac{1}{2}^-)$ & 150 & 0 & 0.9 & 4 \\
N(1680)  & $\frac{1}{2}( \frac{5}{2}^+)$ & 130 & 1&  6.3 & 6/5 \\
N(1720)  & $\frac{1}{2}( \frac{3}{2}^+)$ & 150 & 1 & 7.8 & 8/3 \\
$\Delta$(1232)  & $\frac{3}{2}( \frac{3}{2}^+)$ & 120 & 1 & 15.3 & 16/9 \\
$\Delta$(1620)  & $\frac{3}{2}( \frac{1}{2}^-)$ & 150 & 0 &2.5 & 8/3 \\
$\Delta$(1700)  & $\frac{3}{2}( \frac{3}{2}^-)$ & 300 & 0 & 5.0 & 16/9 \\
$\Delta$(1905)  & $\frac{3}{2}( \frac{5}{2}^+)$ & 350 & 1 & 12.2 & 4/5
\end{tabular}
\end{center}
\label{tab}

\end{table}

\begin{table}
\caption{Parameters used in a consistent model for the
selected subset of resonances.}

\begin{center}
\begin{tabular}{cccc}
 & $I(J^P)$ & $\Gamma_{Total}$(GeV) & $f_{RN\rho}$   \\
\hline\hline
N(940)  & $\frac{1}{2}( \frac{1}{2}^+)$ & 0 & 8.1  \\
N(1520)  & $\frac{1}{2}( \frac{3}{2}^-)$ & 120 & 7  \\
N(1680)  & $\frac{1}{2}( \frac{5}{2}^+)$ & 130 & 6.3  \\
N(1720)  & $\frac{1}{2}( \frac{3}{2}^+)$ & 150 & 7.2  \\
$\Delta$(1232)  & $\frac{3}{2}( \frac{3}{2}^+)$ & 120 & 16 \\
$\Delta$(1905)  & $\frac{3}{2}( \frac{5}{2}^+)$ & 350 & 9.0
\end{tabular}
\end{center}
\label{tab2}

\end{table}

\begin{figure}[htb]
\begin{minipage}[t]{77mm}
\vbox to 2.3in{\vss
   \hbox to 1.5in{\includegraphics{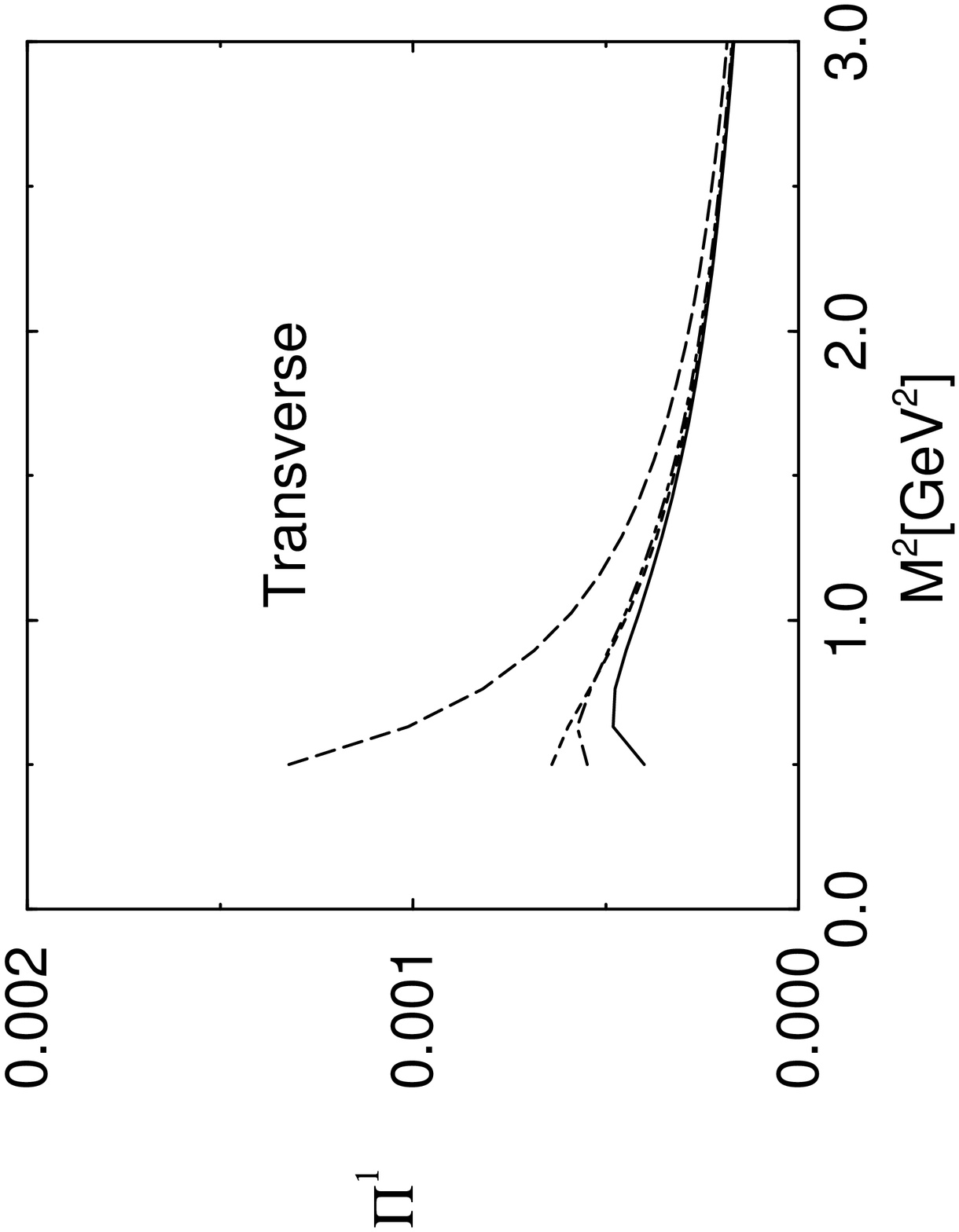}\hss}}
\end{minipage}
\hspace{\fill}
\begin{minipage}[t]{77mm}
\vbox to 2.3in{\vss
   \hbox to 1.5in{\includegraphics{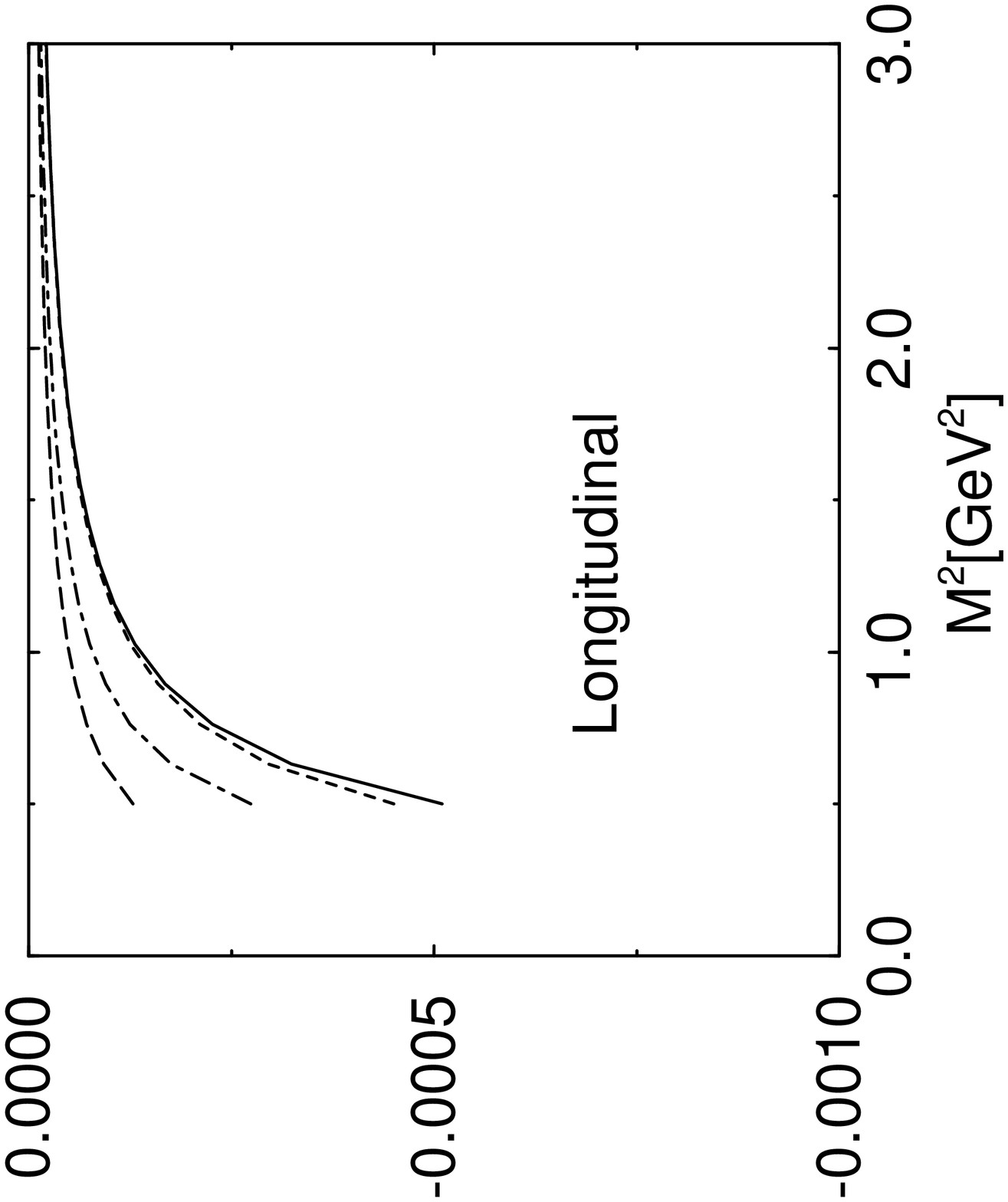}\hss}}
\label{fig:l1}
\end{minipage}
\caption{The OPE in GeV${}^{-2}$. The solid line shows the full calculation,
while the short-dashed line includes the infinite sum of the twist-2
operators. The dashed line shows the OPE without dim-6 and the dash-dotted
without twist-4 operators. All curves are at nuclear matter density.}
\end{figure}
\begin{figure}
\vbox to 3.3in{\vss
   \hbox to 2.5in{\includegraphics{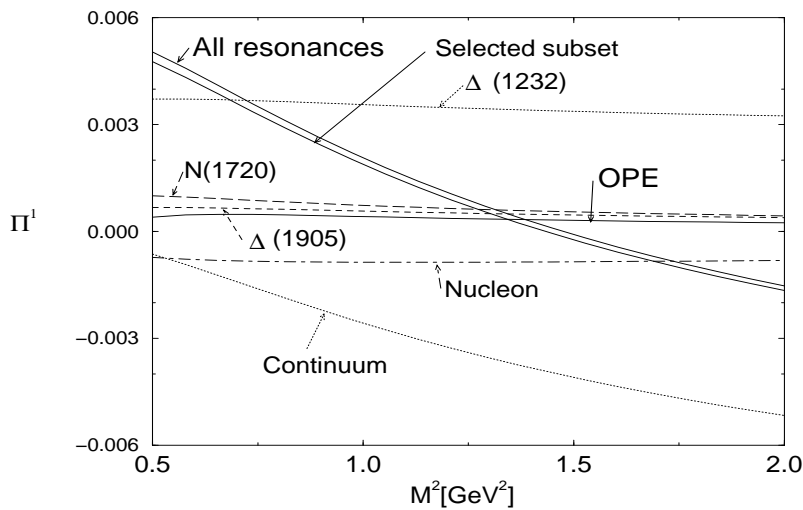}\hss}}
\caption{Real part of $\Pi^1$ in GeV$^{-2}$.  The solid lines denote the
OPE and the best fit including all resonances in the model of ref.~[31].
The contributions from the nucleon, $N(1720),\Delta(1232), \Delta(1905)$
and the continuum are also shown. Furthermore, the contribution of the
most important resonances (the selected subset) is shown for the parameters
given in Table II.}
\end{figure}
\begin{figure}
\vbox to 3.3in{\vss
   \hbox to 2.5in{\includegraphics{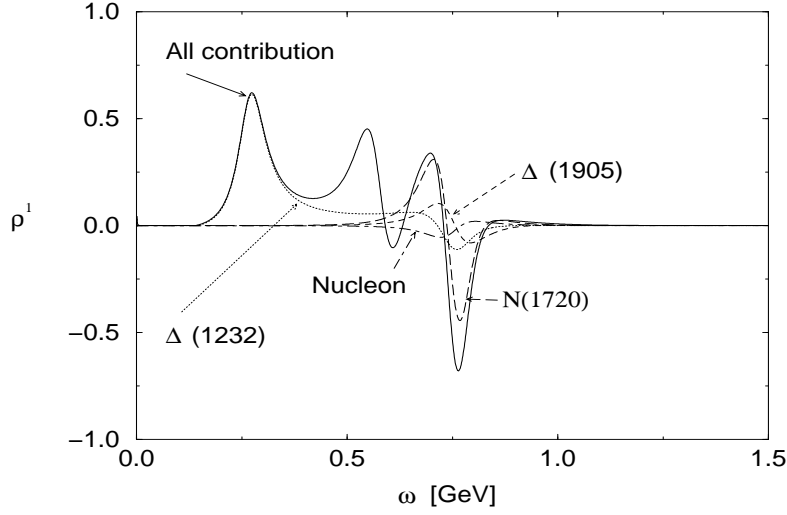}\hss}}
\label{fig:3}
\caption{Imaginary part of $\Pi^1=\rho^1$ in GeV$^{-2}$ as a function of
the energy $\omega$.}
\end{figure}

\begin{figure}
\vbox to 3.3in{\vss
   \hbox to 2.5in{\includegraphics{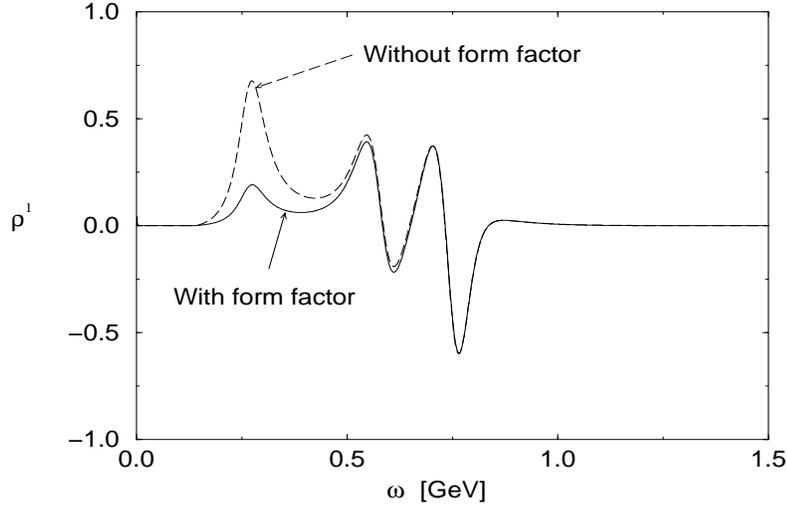}\hss}}
\caption{Imaginary part of $\Pi^1=\rho^1$ in GeV$^{-2}$ for the selected
subset (Table II)  with and without form factors (38) for the nucleon and
the $\Delta(1232)$. }
\end{figure}
\begin{figure}
\vbox to 3.3in{\vss
   \hbox to 2.5in{\includegraphics{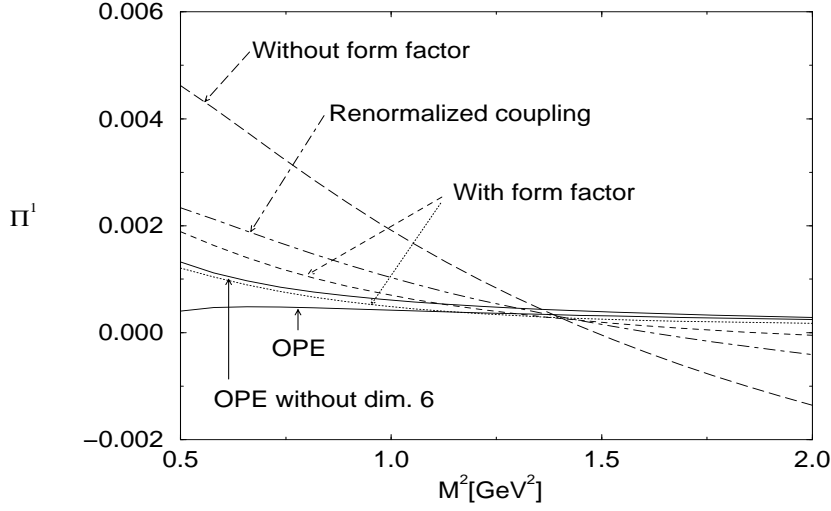}\hss}}
\caption{Real part of $\Pi^1$ in GeV$^{-2}$ with the parameters
given in Table III. The long-dashed line shows the phenomenological side
without the form factor for the nucleon and the delta, while the dashed
line shows the result including the form factors (38). The dotted line
is obtained with the form factors but with zero coupling constants for
the $N(1520)$ and $N(1680)$ resonances. Finally, the dash-dotted line
shows the phenomenological side obtained when the frequency dependence of
the form factors (38) is neglected.}
\label{fig5}
\end{figure}

\end{document}